\documentclass[journal]{IEEEtran}
\usepackage{cite}
\usepackage{amsmath,amssymb,amsfonts}
\usepackage{graphicx}
\usepackage{textcomp}
\usepackage{xcolor}
\usepackage{hyperref}
\hypersetup{hidelinks=true}
\usepackage{algorithm,algorithmic}
\usepackage{tikz}
\usetikzlibrary{positioning, arrows.meta, shapes.geometric, calc, fit}
\def\BibTeX{{\rm B\kern-.05em{\sc i\kern-.025em b}\kern-.08em
    T\kern-.1667em\lower.7ex\hbox{E}\kern-.125emX}}
\newcommand{\ieeepreprintnote}{%
  \begingroup
  \renewcommand{\thefootnote}{}
  \footnotetext{This work has been submitted to the IEEE for possible publication. 
  Copyright may be transferred without notice, after which this version may no longer be accessible.}%
  \addtocounter{footnote}{-1}
  \endgroup
}

\title{A Unified Transformer Architecture for Low-Latency and Scalable Wireless Signal Processing}
\author{%
Yuto Kawai,~Rajeev Koodli\\
Research Institute of Advanced Technology, SoftBank Corp., Tokyo, Japan\\
\textit{Corresponding author:} yuto.kawai@g.softbank.co.jp
}

\begin{document}
\maketitle
\ieeepreprintnote

\begin{abstract}
We propose a unified Transformer-based architecture for wireless signal processing tasks, offering a low-latency, task-adaptive alternative to conventional receiver pipelines. Unlike traditional modular designs, our model integrates channel estimation, interpolation, and demapping into a single, compact attention-driven architecture designed for real-time deployment.
The model's structure allows dynamic adaptation to diverse output formats by simply modifying the final projection layer, enabling consistent reuse across receiver subsystems. Experimental results demonstrate strong generalization to varying user counts, modulation schemes, and pilot configurations, while satisfying latency constraints imposed by practical systems.
The architecture is evaluated across three core use cases: (1) an End-to-End Receiver, which replaces the entire baseband processing pipeline from pilot symbols to bit-level decisions; (2) Channel Frequency Interpolation, implemented and tested within a 3GPP-compliant OAI+Aerial system; and (3) Channel Estimation, where the model infers full-band channel responses from sparse pilot observations. In all cases, our approach outperforms classical baselines in terms of accuracy, robustness, and computational efficiency.
This work presents a deployable, data-driven alternative to hand-engineered PHY-layer blocks, and lays the foundation for intelligent, software-defined signal processing in next-generation wireless communication systems.
\end{abstract}


\maketitle

\section{INTRODUCTION}
\label{sec:introduction}

\IEEEPARstart{W}{ireless} communication systems are undergoing a paradigm shift driven by surging mobile traffic, the diversification of service requirements, and the growing complexity of deployment scenarios. Emerging applications -- ranging from mission-critical systems requiring near-instantaneous response and flawless reliability to large-scale sensor networks and immersive extended reality (XR) -- demand not only high throughput and low latency, but also adaptability to dynamic and heterogeneous channel conditions. These trends challenge the design principles of traditional receivers, which typically rely on cascaded signal processing blocks—synchronization, channel estimation, equalization, demapping, and decoding—implemented using deterministic algorithms with hand-crafted heuristics and strong domain assumptions~\cite{Dai2020,Zhang2019}.

While this modular design has served well in static or moderately dynamic environments, its limitations become pronounced in modern systems such as 5G NR and beyond, where flexible numerologies, MU-MIMO, dynamic scheduling, and interference-limited deployments are commonplace. In such systems, blockwise optimization may yield suboptimal end-to-end performance, and rigid assumptions on pilot patterns or fading statistics often lead to brittleness when conditions deviate from the design-time assumptions. Moreover, implementing and tuning each module independently complicates the design pipeline, making it costly to adapt to new spectrum bands, carrier aggregations, or hardware impairments.

Over the past few years, deep learning (DL) has emerged as a compelling alternative to this paradigm~\cite{Ye2018,He2018,Samuel2019,Shlezinger2021,Dorner2018,Felix2018,OShea2017,Letizia2021,Raviv2023}. Neural networks can approximate complex nonlinear mappings directly from data, enabling them to learn statistical patterns that are difficult to model analytically. In the physical layer, DL-based approaches have been investigated for tasks such as channel estimation, equalization, decoding, and even end-to-end learned communication systems that jointly optimize transmitter and receiver design.

From an architectural perspective, convolutional neural networks (CNNs) have been widely used due to their ability to exploit local time-frequency structures, but they are often constrained by limited receptive fields and fixed architectural configurations. In contrast, the Transformer architecture~\cite{vaswani2017attention} leverages self-attention to capture long-range dependencies and structured correlations across time, frequency, and spatial dimensions, offering greater adaptability to varying system parameters~\cite{Cai2022TRNAMC}. These properties align well with the characteristics of wireless channels, which exhibit structured correlations due to Doppler spread, multipath delay, and pilot placement.

Nevertheless, most existing Transformer-based studies in wireless remain at the algorithmic or simulation stage, without fully addressing the constraints of real-time deployment—such as latency budgets, memory and power limits, and integration into standardized wireless stacks~\cite{Doha2025DLReceiverSurvey,Jiao2021ChinaComms}. \textit{This paper directly addresses this gap} by proposing a Transformer-based architecture specifically tailored for wireless signal processing. Unlike generic adaptations from language or vision tasks, our design operates directly on resource elements, incorporates OFDM-aligned positional encodings, and preserves signal amplitude in early layers. The architecture is compact, modular, and adaptable to multiple PHY-layer tasks without redesign.

We validate the proposed model in three representative use cases, each highlighting distinct aspects of practical deployment and architectural flexibility:
\begin{itemize}
  \item \textbf{End-to-End Receiver:} Drop-in replacement for the channel estimation to demapping chain, adaptable to various antenna settings, modulation schemes, and user densities.
  \item \textbf{Channel Frequency Interpolation:} Real-time over-the-air validation using a 3GPP-compliant stack~\cite{NVIDIA_ARCOTA_ProductDesc_2025}, emphasizing deployment readiness under practical interference conditions.
  \item \textbf{Channel Estimation:} Modular PHY-layer evaluation without architectural changes, illustrating the versatility of the shared Transformer backbone.
\end{itemize}

Together, these results show that a unified attention-based backbone can serve as a building block for next-generation AI-native receivers, supporting both fine-grained inference for individual processing blocks and complete end-to-end operation.

The remainder of this paper is organized as follows: Section~\ref{sec:related_work} surveys related work on deep learning for physical layer design in greater detail. Section~\ref{sec:architecture} presents the proposed Transformer architecture and its signal-domain adaptations. Section~\ref{sec:use_cases} details the implementation of each use case and corresponding evaluations. Section~\ref{sec:discussion} discusses deployment trade-offs, scalability, and interpretability. Section~\ref{sec:future_work} outlines future research directions, and Section~\ref{sec:conclusion} concludes the paper.

\section{RELATED WORK}
\label{sec:related_work}

Deep learning (DL) methods have rapidly gained traction in wireless communication, offering a data-driven alternative to conventional block-based physical layer design~\cite{Dai2020,Zhang2019}. Early research primarily focused on replacing individual components in the receiver pipeline, such as channel estimation~\cite{Ye2018,He2018,Soltani2019}, equalization~\cite{Samuel2019,Shlezinger2021}, and decoding~\cite{Dorner2018,Felix2018}. These systems have often utilized convolutional neural networks (CNNs), leveraging their ability to capture local spatial and spectral correlations in time-frequency signal representations. Notably, DeepRx~\cite{honkala2021deeprx} demonstrated that fully convolutional neural receivers could replace traditional estimation and detection blocks with a unified architecture, achieving improved robustness and lower latency than LMMSE baselines in OFDM systems.

To handle the increasing complexity of MU-MIMO systems and flexible pilot allocations in 5G NR, Cammerer et al.~\cite{cammerer2023neuralreceiver} proposed a hybrid CNN-GNN-based end-to-end receiver. Their model demonstrated robust performance across user counts and pilot patterns, highlighting the benefits of integrating geometric structure into neural architectures. Simultaneously, a parallel stream of research explored end-to-end learning through autoencoder frameworks~\cite{OShea2017,Felix2018,Dorner2018}. These systems jointly optimized transmitters and receivers for specific channels, enabling customized modulation schemes and demonstrating gains over classical designs in nonlinear or hardware-impaired environments~\cite{Letizia2021,Raviv2023}.

However, early CNN-based or autoencoder-based models often struggled to capture global dependencies across wideband signals, long symbol durations, or multi-antenna spatial structures. To overcome these limitations, researchers began applying Transformer architectures—originally developed for NLP—to wireless signal processing tasks. The self-attention mechanism within Transformers provides a natural way to model inter-symbol and inter-carrier relationships, making them well-suited to wireless environments characterized by multipath fading, interference, and nonstationarity.

Channelformer~\cite{luan2022channelformer} and SigT~\cite{ren2022sigt} were among the first to apply Transformers to channel estimation and joint equalization tasks, respectively. These models achieved improved accuracy and generalization across channel models compared to CNN baselines. Comm-Transformer~\cite{xie2024commtransformer} extended the framework to frequency-domain input and demonstrated that attention mechanisms can outperform classical interpolation even with sparse pilots. Additional work by Leng et al.~\cite{leng2025complextransformer} introduced complex-valued Transformers that respect the structure of I/Q data, improving estimation and detection in phase-sensitive systems.

Semantic communication has also been explored with Transformer variants, where Zhou et al.~\cite{zhou2021semantic} proposed an adaptive Universal Transformer to handle varying semantic complexity and channel conditions.

Transformers have also been explored in mmWave and OTFS systems~\cite{sun2023,Gao2022}, where they model delay-Doppler sparsity and angular resolution. Hybrid architectures combining CNN frontends with Transformer attention backends~\cite{yue2024,Yang2021} have been proposed as latency-efficient alternatives, leveraging local and global feature extraction in tandem. 

Recently, Khawaja et al.~\cite{khawaja2024lwm} proposed the Large Wireless Model (LWM), a Transformer-based foundation model trained using masked channel modeling over a large-scale, diverse wireless dataset. LWM focuses on learning general-purpose channel embeddings that can be reused across downstream tasks, showing strong performance in representation learning. However, it does not target deployment within latency-sensitive baseband receivers, nor does it evaluate real-time or over-the-air (OTA) integration.

Despite these promising results, most studies evaluate models in simulation, without assessing real-time viability or compliance with 3GPP-based system constraints.

Several challenges hinder practical deployment of Transformer-based receivers. 
The quadratic complexity of self-attention poses issues for long OFDM symbols, large bandwidths, or massive MIMO arrays~\cite{Tay2022Efficient,Dao2023Flash2,Dao2022Flash}. 
Moreover, integrating these models into real-world wireless stacks—such as those governed by 3GPP standards—requires hardware-aware design and improved interpretability. 
Recent surveys~\cite{Dai2020,Doha2025DLReceiverSurvey,Jiao2021ChinaComms} and the 3GPP Release-18 study on AI/ML for the NR air interface~\cite{TR38843,LinArxiv2023,LinCTN2025} 
highlight the potential of AI/ML at the PHY and air-interface, while noting that further work is needed toward robust deployment, testing, and reliability at scale.

To address these limitations, we propose a unified Transformer architecture tailored for wireless signal processing. Unlike task-specific designs, our model is compact, modular, and validated across both simulation and real-time OTA testing. Through three use cases -- End-to-End Receiver, Channel Frequency Interpolation, and Channel Estimation -- we demonstrate the model's ability to match or surpass classical algorithms while satisfying practical deployment constraints such as latency, throughput, and task generality.

\section{PROPOSED TRANSFORMER ARCHITECTURE}
\label{sec:architecture}

The proposed Transformer-based architecture is a compact and latency-aware model tailored for wireless signal processing. Unlike traditional receivers, which rely on sequential and domain-specific signal processing blocks such as channel estimation, equalization, and demapping, our architecture unifies these operations into a single data-driven model. Its design emphasizes practical deployability, offering both high accuracy and computational efficiency across a wide range of tasks—including end-to-end receiver, channel frequency interpolation, and channel estimation.

While the model incorporates standard Transformer components such as self-attention and feedforward layers, its overall design is specialized for wireless signal processing. In contrast to general-purpose Transformer architectures, our model processes signals at the granularity of individual resource elements and omits architectural components—such as early-stage normalization or patch-based tokenization—that are not well suited to the characteristics of physical-layer data. Instead, the architecture reflects the unique demands of wireless systems, where absolute signal amplitude, per-element resolution, and real-time inference capability are critical. This results in a task-adaptive yet computationally efficient model suitable for deployment under tight latency constraints, such as those required for feedback-driven processes in 5G and beyond.

\subsection{ARCHITECTURE OVERVIEW}
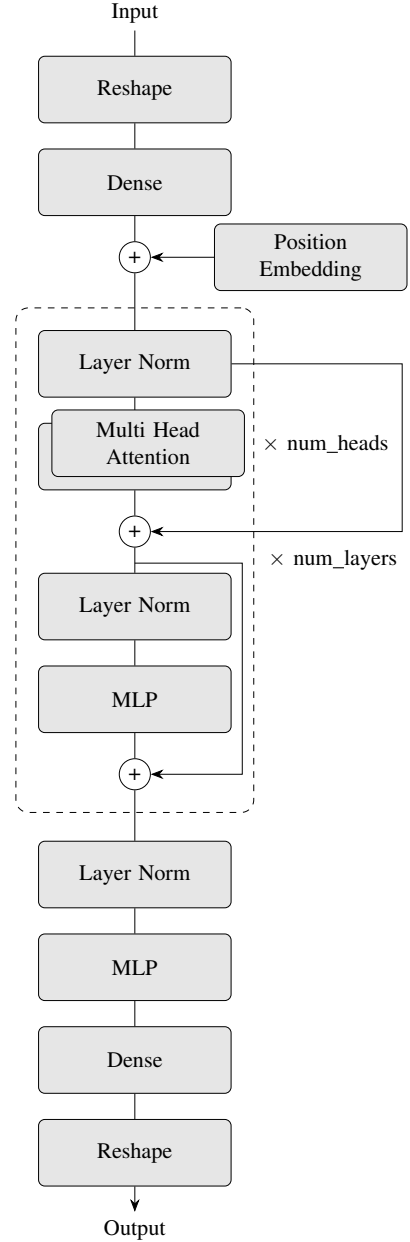
\begin{figure}[t]
    \centering
    \resizebox{0.6\linewidth}{!}{\begin{tikzpicture}[
  block/.style={rectangle, draw, fill=gray!20,
                text width=8em, text centered,
                minimum height=3em, rounded corners=1mm},
  line/.style ={draw,-{Stealth[length=2mm]}},
  sum/.style  ={circle, draw, minimum size=1.4em, label=center:+},
  box/.style  ={draw,dashed,inner sep=10pt,rounded corners=2mm}
]

\newcommand{\vspacegap}{0.4cm}   % 通常の上下間隔
\newcommand{\firstgap}{0.9cm}    % Position Embedding 直下だけ広げる
\newcommand{\afterEncoder}{0.8cm}% ← sum_ff と次の LN の余白

%-----------------------------------------------------------------
% ノード配置
%-----------------------------------------------------------------
\node (input)                              {Input};
\node[block,below=\vspacegap of input]     (reshape_in)  {Reshape};
\node[block,below=\vspacegap of reshape_in](dense_in)    {Dense};
\node[sum,  below=\vspacegap of dense_in]  (sum_pos)     {};

% Position Embedding
\node[block,right=1.0cm of sum_pos]        (pos_embed) {Position\\Embedding};
\draw[line] (pos_embed.west) -- (sum_pos.east);

% エンコーダ内部
\node[block,below=\firstgap of sum_pos]    (ln_core1)   {Layer Norm};
\node[block,below=\vspacegap of ln_core1,
      name=mha_back,text opacity=0] {};
\node[sum,  below=\vspacegap of mha_back]  (sum_attn)   {};
\node[block,below=\vspacegap of sum_attn]  (ln_core2)   {Layer Norm};
\node[block,below=\vspacegap of ln_core2]  (mlp_core)   {MLP};
\node[sum,  below=\vspacegap of mlp_core]  (sum_ff)     {};

% エンコーダ外（余白を afterEncoder）
\node[block,below=\afterEncoder of sum_ff] (ln_out)     {Layer Norm};
\node[block,below=\vspacegap of ln_out]    (mlp_out)    {MLP};
\node[block,below=\vspacegap of mlp_out]   (dense_out)  {Dense};
\node[block,below=\vspacegap of dense_out] (reshape_out){Reshape};
\node[below=\vspacegap of reshape_out]     (output)     {Output};

%-----------------------------------------------------------------
% 主経路
%-----------------------------------------------------------------
\path[line] (input) -- (reshape_in)
            (reshape_in) -- (dense_in)
            (dense_in) -- (sum_pos)
            (sum_pos) -- (ln_core1)
            (ln_core1) -- (mha_back)
            (mha_back) -- (sum_attn)
            (sum_attn) -- (ln_core2)
            (ln_core2) -- (mlp_core)
            (mlp_core) -- (sum_ff)
            (sum_ff) -- (ln_out)
            (ln_out) -- (mlp_out)
            (mlp_out) -- (dense_out)
            (dense_out) -- (reshape_out)
            (reshape_out) -- (output);

%-----------------------------------------------------------------
% 残差線（枠内に収める）
%-----------------------------------------------------------------
\def\skipone{2.7cm}
\def\skiptwo{1.7cm}

\draw[line] (ln_core1.east) -- ++(\skipone,0) |- (sum_attn.east);
\coordinate (skip2start) at ($(sum_attn.center)+(0,-0.5)$);
\draw[line] (skip2start) -- ++(\skiptwo,0) |- (sum_ff.east);

%-----------------------------------------------------------------
% Encoder 点線枠（Layer Norm～sum_ff を囲む）
%-----------------------------------------------------------------
\node[box,fit={(ln_core1) (sum_ff)}] (encbox) {};
\node[right=0.1cm of encbox.east] {$\times$ num\_layers};

%-----------------------------------------------------------------
% MHA 前面ブロック（矢印を隠す装飾）
%-----------------------------------------------------------------
\node[block,anchor=center,name=mha_front,
      at={($(mha_back.center)+(0.6em,0.6em)$)}]%
      {Multi Head Attention};
\node[right=0.15cm of mha_front.east] {$\times$ num\_heads};

\end{tikzpicture}}
    \caption{Overview of the proposed latency-aware Transformer architecture. The model operates on resource-element-level embeddings, omits early normalization, and supports multiple PHY-layer tasks via task-specific output heads.}
    \label{fig:architecture}
\end{figure}

The input to the model consists of time-frequency structured tensors, optionally including spatial dimensions such as antenna indices as illustrated in Figure~\ref{fig:architecture}. These are flattened along the time and frequency axes to form a sequence of resource elements (REs), which represent individual OFDM symbols and subcarriers. A Dense projection is applied independently to each RE, embedding the associated features from other dimensions (e.g., antennas) into a shared feature space. This enables the model to treat each RE as an independent token while preserving the resolution necessary for per-element predictions.

A positional encoding layer is then applied to inject time and frequency context, allowing the attention mechanism to learn dependencies across REs. Unlike typical Transformers, we omit Layer Normalization in early stages to preserve signal magnitude, which is essential for tasks involving channel estimation and physical-layer reconstruction.

The encoded RE tokens are passed through one or more layers of Multi-Head Self-Attention (MHSA) followed by feedforward networks. These layers model correlations across both the frequency and time dimensions, capturing effects such as multipath propagation and inter-symbol interference. Importantly, experimental results show that excessively increasing the number of attention heads does not improve performance, allowing us to maintain architectural simplicity.

Following the Transformer encoder, a lightweight post-processing stack composed of Layer Normalization, an MLP, and a final Dense projection produces the task-specific output. These operations are applied independently across REs, ensuring that the model can be used for both element-wise regression (e.g., channel estimation) and classification (e.g., bit decoding) tasks.

\paragraph*{Default instantiation}
Unless otherwise noted, we instantiate the unified Transformer with four encoder layers and four attention heads per layer.
This serves as a consistent reference configuration across tasks, chosen for its generally strong performance in our preliminary experiments rather than as the result of task-specific optimization.
In certain cases -- such as Channel Frequency Interpolation -- lighter variants (e.g., a single encoder layer) are used, and such deviations are stated in the corresponding task-specific sections.

\subsection{TASK ADAPTABILITY AND DEPLOYMENT}
A key feature of the proposed architecture is its ability to support diverse wireless signal processing tasks using a shared backbone. This is enabled by modifying only the final output projection and selecting an appropriate loss function, without altering the core model structure. Specifically:

\begin{itemize}
  \item \textbf{End-to-End Receiver:} The model outputs soft bits or bit probabilities. It is trained using binary cross-entropy (BCE) loss, targeting bit-level decisions after equalization and demapping.

  \item \textbf{Channel Frequency Interpolation:} The model predicts complex-valued channel coefficients at unobserved frequency positions. It uses mean squared error (MSE) loss computed between the predicted and ground-truth channel values.

  \item \textbf{Channel Estimation:} The model reconstructs the full channel response across time and frequency. This regression task also uses MSE loss, applied to all subcarriers and OFDM symbols within a resource block.
\end{itemize}

Despite these different output targets, the underlying architecture remains unchanged. This design enables both modular deployment—where the model replaces a specific processing stage—and end-to-end integration—where it subsumes multiple blocks such as channel estimation and demapping.

Training is performed in a supervised manner using labeled datasets derived from simulated or real over-the-air signals. To enhance generalization across deployment scenarios, we apply data augmentation strategies such as varying SNR levels, channel models, and antenna permutations.

The model is also optimized for deployment in latency-sensitive environments. Its lightweight structure -- with limited attention heads and shallow depth -- achieves sub-millisecond PUSCH pipeline execution in real-time OTA evaluations using the OAI+Aerial platform~\cite{NVIDIA_ARCOTA_ProductDesc_2025}, meeting the timing requirements for HARQ feedback and dynamic scheduling in 5G NR systems.

\section{EMPIRICAL VALIDATION}
\label{sec:use_cases}

We validate the proposed Transformer architecture through three representative use cases, each targeting a distinct aspect of wireless signal processing and evaluated under different environments:

\begin{itemize}
  \item \textbf{End-to-End Receiver (Simulation):} Demonstrates end-to-end capability by replacing the entire receiver chain from channel estimation to demapping. Evaluated via link-level simulations using the Sionna framework with 3GPP-compliant CDL-C channels in both SIMO and MU-MIMO scenarios, highlighting adaptability to user count, pilot structure, and modulation scheme.

  \item \textbf{Channel Frequency Interpolation (OTA):} Validates real-time inference performance through over-the-air (OTA) evaluation using the OAI+Aerial stack under measured channel conditions. This task emphasizes deployment feasibility under strict latency and system integration constraints.

  \item \textbf{Channel Estimation (Simulation):} Serves as a standalone regression task to showcase the model's versatility and applicability to conventional PHY-layer modules without architectural modification. Evaluated via Sionna-based simulations with 3GPP-compliant CDL-C channels.
\end{itemize}

All use cases share the same core Transformer backbone. Only the output projection and loss function are adapted for each task. This design demonstrates that a single model can support diverse processing roles—from bit-level decisions to channel reconstruction—across both simulation and real-world OTA environments without retraining the entire system.

The following subsections detail each task’s configuration, objective, and evaluation, providing a comprehensive assessment of the model’s flexibility, efficiency, and performance across deployment contexts.

\subsection{END-TO-END RECEIVER}
\label{sec:neural_receiver}

This task evaluates the capability of the proposed Transformer to operate as a full-stack end-to-end receiver, replacing the entire traditional receiver chain from channel estimation to demapping and bit recovery. The model directly infers log-likelihood ratios (LLRs) for each transmitted bit per resource element (RE), enabling complete end-to-end decoding from the raw received signal without requiring intermediate blocks.

We investigate two primary experimental settings:

\begin{enumerate}
  \item \textbf{Single-user (1UE), pilot-fixed}: A controlled setting for direct comparison against existing baseline models (e.g., CNN-based, LS+MMSE).
  \item \textbf{Multi-user (2UE), pilot-random}: A practical MU-MIMO configuration evaluating the model's robustness against inter-user interference and scalability.
\end{enumerate}

In both settings, the model shares the same Transformer architecture. The input is processed in non-overlapping tiles of size \texttt{12 subcarriers × 14 OFDM symbols}, ensuring scalability to arbitrary frame durations. The output dimension is determined by the number of users and modulation order, and the model is trained to produce all LLR bits simultaneously. This allows flexible decoding for any modulation format (e.g., QPSK, 64-QAM) by selecting only relevant bits—enabling dynamic adaptation to varying MCS without retraining.

\subsubsection{Experimental Setup}

The evaluation uses the Sionna framework with 3GPP-compliant CDL-C channels. Table~\ref{tab:neural_receiver_params} summarizes key parameters for simulation.

\begin{table}[t]
  \centering
  \caption{Simulation Parameters for End-to-End Receiver Evaluation}
  \label{tab:neural_receiver_params}
  \begin{tabular}{ll}
    \hline
    \textbf{Parameter} & \textbf{Value} \\
    \hline
    Carrier Frequency & 3.5 GHz \\
    Subcarrier Spacing & 30 kHz \\
    OFDM Symbols per Slot & 14 \\
    Subcarriers per RB & 12 \\
    Modulation Schemes & 256-QAM \\
    Antenna Configuration & Uplink: 1 or 2 UEs (1 Tx each), BS: 4 Rx \\
    Channel Model & CDL-C \\
    $E_b/N_0$ Range & 0--40 dB \\
    Training Samples & 100k \\
    Loss Function & Binary Cross-Entropy (BCE) \\
    Evaluation Framework & Sionna (TensorFlow) \\
    \hline
  \end{tabular}
\end{table}

We compare our Transformer-based End-to-End Receiver against a range of baselines:

\begin{itemize}
  \item \textbf{CNN-based End-to-End Receiver}: A single-user model provided as a reference implementation by the Sionna framework\footnote{\url{https://github.com/NVlabs/sionna}}. To ensure a fair comparison, we use the publicly available version without architectural modifications.
  \item \textbf{Perfect CSI}: Idealized reference using ground-truth channel coefficients, combined with MMSE detection.
  \item \textbf{LS Estimation + Linear Interpolation}: Channel estimation is performed at pilot positions via least squares (LS), followed by two-dimensional linear interpolation in time and frequency.
  \item \textbf{LS Estimation + Nearest Neighbor Interpolation}: Non-pilot positions are assigned the value of the nearest DMRS observation.
\end{itemize}

Note that the CNN baseline is evaluated only in the single-user setting, as the provided reference does not natively support multi-user processing. While extension to MU-MIMO is technically feasible, we restrict our evaluation to publicly released implementations to ensure reproducibility and fairness.

\subsubsection{Experimental Results}

\paragraph{1UE: Pilot-Fixed Baseline Comparison}

\begin{figure}[t]
  \centering
  \includegraphics[width=1\linewidth]{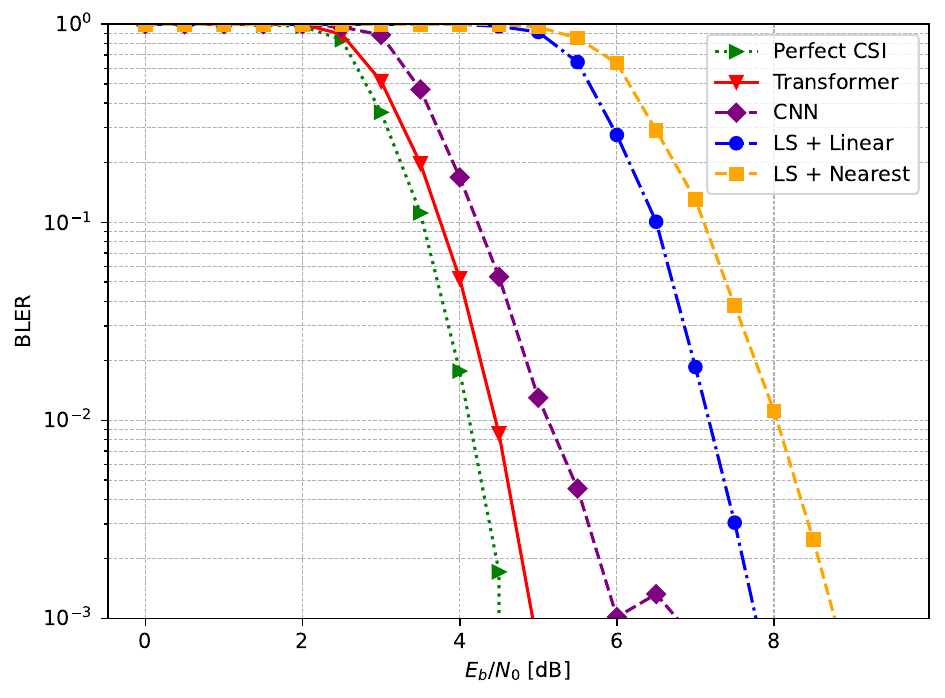}
  \caption{BLER in single-user SIMO (1UE, 4Rx). Transformer vs CNN vs LS vs Perfect CSI.}
  \label{fig:neural_receiver_simo}
\end{figure}

Figure~\ref{fig:neural_receiver_simo} shows the BLER under a single-user SIMO setting. The Transformer significantly outperforms the LS-based baseline and surpasses the CNN-based model, approaching the perfect CSI reference.

\paragraph{2UE: Pilot-Random MU-MIMO Evaluation}

\begin{figure}[t]
  \centering
  \includegraphics[width=1\linewidth]{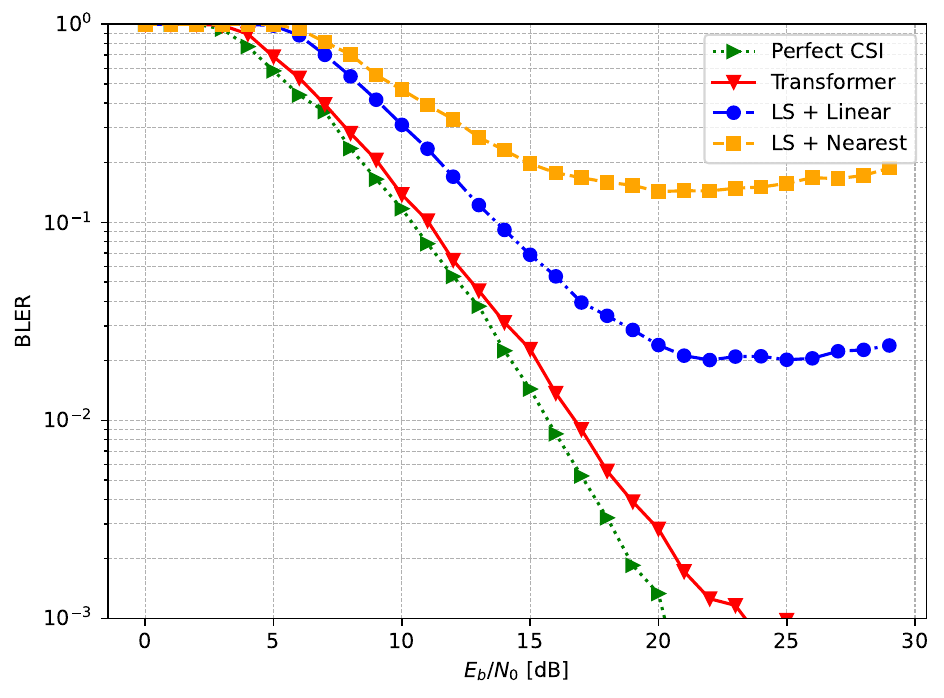}
  \caption{BLER in 2-user MU-MIMO: Transformer vs LS and Perfect CSI.}
  \label{fig:neural_receiver_mimo_baseline}
\end{figure}

Figure~\ref{fig:neural_receiver_mimo_baseline} presents BLER performance for a 2-user MU-MIMO setup. Without any dedicated separation modules, the Transformer still achieves strong performance close to the perfect CSI baseline and significantly outperforms the conventional LS+MMSE pipeline.

\paragraph{2UE: Architectural Sensitivity (Layer Depth)}

\begin{figure}[t]
  \centering
  \includegraphics[width=1\linewidth]{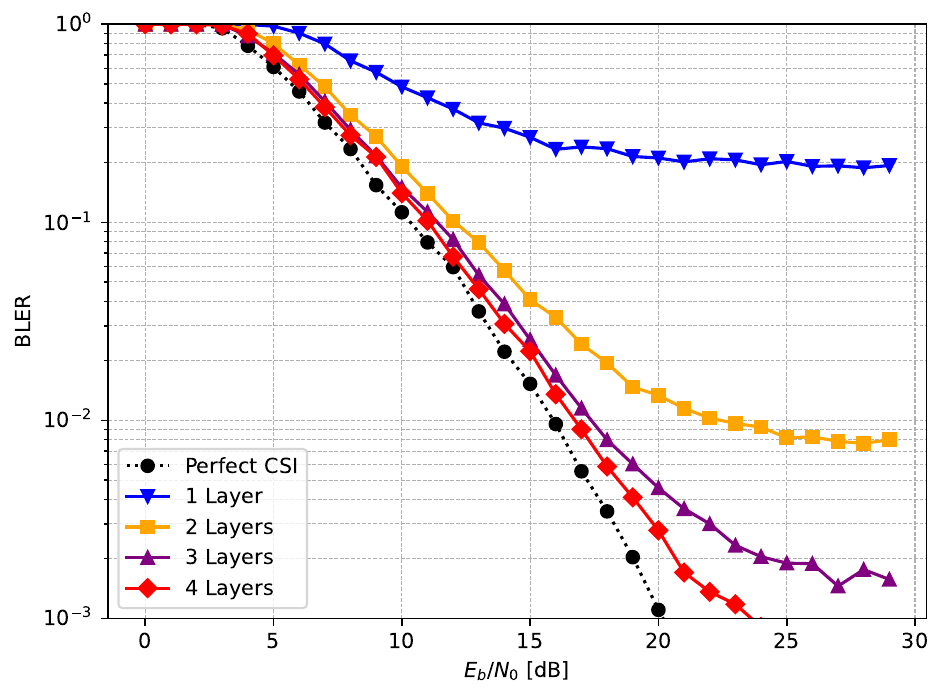}
  \caption{BLER performance with varying Transformer depth (1–4 layers) in 2-user MU-MIMO.}
  \label{fig:neural_receiver_layer_sweep}
\end{figure}

To evaluate the impact of model depth, Figure~\ref{fig:neural_receiver_layer_sweep} presents the BLER performance for Transformer models with 1 to 4 layers, under fixed head count in a 2-user MU-MIMO scenario. Performance improves notably from 1 to 2 layers, but further increases in depth yield only marginal gains. The 2-layer model already approaches the Perfect CSI bound, and the 4-layer model saturates performance. These results highlight a key strength of the proposed architecture: high performance can be achieved with shallow, low-latency models—an essential feature for practical deployment in real-time systems.

\paragraph{2UE: Architectural Sensitivity (Attention Heads)}

\begin{figure}[t]
  \centering
  \includegraphics[width=1\linewidth]{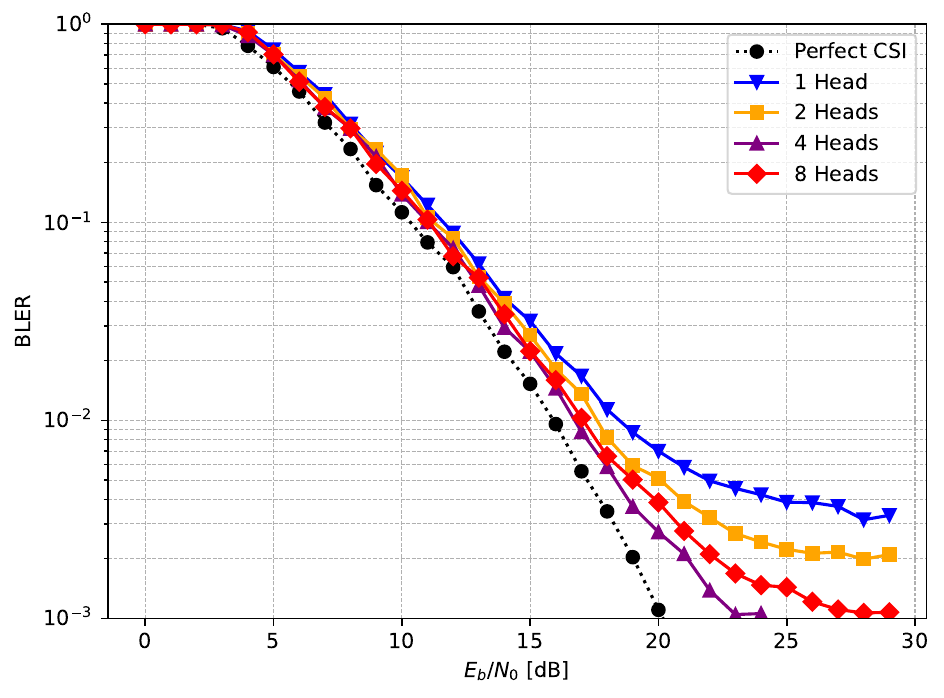}
  \caption{BLER performance with varying attention head count (1–8 heads) in 2-user MU-MIMO.}
  \label{fig:neural_receiver_head_sweep}
\end{figure}

Figure~\ref{fig:neural_receiver_head_sweep} shows the BLER curves for different attention head counts (1–8 heads), with layer depth fixed. The model exhibits stable performance across this range, and even a single-head configuration performs comparably to multi-head variants. This robustness suggests that high head counts are not essential, enabling further reduction in complexity and latency without sacrificing accuracy.

Together, these findings demonstrate that the proposed Transformer architecture sustains high performance under realistic MU-MIMO conditions while being remarkably tolerant to architectural simplification. The ability to operate effectively with shallow depth and minimal attention heads makes it a highly practical choice for scalable, low-latency wireless deployments.

\subsection{CHANNEL FREQUENCY INTERPOLATION}
\label{sec:channel_interp}

We next validate the versatility of our Transformer model in the task of channel frequency interpolation, a task that demands fine-grained estimation accuracy under practical deployment constraints. Unlike End-to-End Receiver or Channel Estimation, which operate on full end-to-end decoding, this task focuses purely on reconstructing missing frequency-domain channel values from sparse pilots, making it a representative low-level signal processing application.

We deploy the model under Over-the-Air (OTA) conditions using the OAI+Aerial framework, where we evaluate real-time inference performance in a SIMO setting (1 UE, 4 BS Rx). The target is to recover the complete channel response across 12 subcarriers and 2 OFDM symbols from sparse DMRS pilots placed in 6 subcarriers and 2 symbols within each resource block. Each input/output is treated as a unit spanning one RB and one slot (12 SC × 14 sym), consistent with 5G NR numerology.

The Transformer model receives a $(6 \times 2 \times 2)$ input tensor—6 pilot subcarriers × 2 symbols × 2 (real and imaginary)—and outputs a $(12 \times 2 \times 2)$ tensor for full-band interpolation over those same 2 OFDM symbols. Since antenna-wise correlations are weak in this task, the four receive antennas are treated as independent samples during training.

\begin{table}[t]
  \centering
  \caption{OTA Parameters for Frequency Interpolation Task}
  \label{tab:interp_params}
  \begin{tabular}{ll}
    \hline
    \textbf{Parameter} & \textbf{Value} \\
    \hline
    Carrier Frequency & 3.73 GHz \\
    Subcarrier Spacing & 30 kHz \\
    DMRS Pattern & $6 \times 2$ (SC $\times$ sym) \\
    Interpolation Target & $12 \times 2$ \\
    Antenna Configuration & 1 UE (1 Tx), 1 BS (4 Rx) \\
    Modulation Scheme & Adaptive (OTA Measured) \\
    Channel Model & OTA Measured \\
    SNR Range & OTA Measured \\
    Training Samples & 220k \\
    Loss Function & MSE \\
    Evaluation Framework & OAI+Aerial \\
    \hline
  \end{tabular}
\end{table}

To prepare OTA-compatible training data, we use a hybrid simulation-capture approach. The transmitted signal $x$ is regenerated using a Sionna PUSCH transmitter, taking as input the captured transport block (TB) extracted from FAPI logs. This allows us to recover the per-resource-element channel as $\hat{h} = y / x$, using fronthaul-captured I/Q samples $y$.

To introduce training robustness, we augment the data by injecting controlled AWGN noise $n$, forming perturbed inputs $y' = (y + n)/x$ with the original $y/x$ as ground-truth. This technique simulates realistic degradation and forces the model to generalize under variable interference conditions.

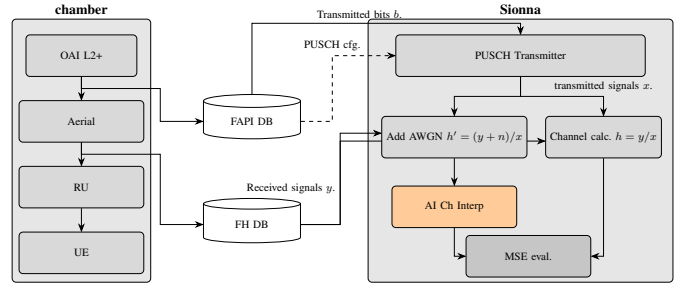
\begin{figure}[t]
    \centering
    \resizebox{1\linewidth}{!}{\begin{tikzpicture}[node distance=6mm and 14mm]

\tikzset{
  block/.style={rectangle, draw, fill=gray!20,
                text centered, font=\footnotesize,
                minimum height=3em, minimum width=9em, rounded corners=1mm},
  ai/.style   ={block, fill=orange!40},
  db/.style   ={cylinder, cylinder uses custom fill,
                cylinder body fill=white, cylinder end fill=white,
                shape border rotate=90, aspect=0.25,
                draw, minimum height=3.2em, minimum width=7em, font=\footnotesize},
  line/.style={draw,-{Stealth[length=2mm]}},
  step/.style={red!70!black, font=\small\bfseries},
}

%================== Chamber block ==================
\node[block, minimum width=10em, minimum height=19em,
      label={[font=\bfseries]above:chamber}] (chamber) {};

% Chamber 内部ノード
\node[block, fill=gray!30, minimum width=8em,
      anchor=north, yshift=-4mm] at (chamber.north) (oai) {OAI L2+};
\node[block, fill=gray!30, below=6mm of oai] (aerial) {Aerial};
\node[block, fill=gray!30, below=6mm of aerial] (ru) {RU};
\node[block, fill=gray!30, below=6mm of ru] (ue) {UE};

%================== Databases (中央) =================
\node[db, right=15mm of aerial] (fapi) {FAPI DB};
\node[db, below=15mm of fapi]       (fh)   {FH DB};

% キャプチャ線
\draw[line] (oai.south) -- (aerial.north);
\draw[line] (aerial.south) -- (ru.north);
\draw[line] (ru.south) -- (ue.north);
\draw[line] (oai.south) |- ++(20mm,-3mm) |- (fapi.west);
\draw[line] (aerial.south) |- ++(20mm,-3mm) |- (fh.west);

%================== Sionna block ==================
\node[block, right=55mm of chamber, minimum width=22em, minimum height=19em,
      label={[font=\bfseries]above:Sionna}] (sionna) {};

% PUSCH TX (全幅)
\node[block, fill=gray!30, minimum width=18em,
      anchor=north, yshift=-4mm] at (sionna.north) (pusch) {PUSCH Transmitter};

% 左: ground-truth path
\node[block, fill=gray!30, minimum width=8em,
      anchor=north west, yshift=-10mm, xshift=38mm] at (pusch.south west) (chan_calc) {Channel calc. $h = y/x$};
\draw[line] (fh.east) -- ++(10mm,0) |- ([yshift=-1mm]chan_calc.west);
% 右: noisy input path
\node[block, fill=gray!30, minimum width=8em,
      anchor=north east, yshift=-10mm, xshift=-30mm] at (pusch.south east) (awgn)    {Add AWGN $h'=(y+n)/x$};
\node[ai, below=7mm of awgn] (ai_ce) {AI Ch Interp};

% MSE at 下中央
\node[block, fill=gray!45, anchor=north] at ($(sionna.south)+(2mm,12mm)$) (mse) {MSE eval.};

% 矢印: PUSCH → Channel calc
\draw[line] (pusch.south)  -- ++(0,-5mm) -| (chan_calc.north)
                        node[midway, above, font=\footnotesize]{transmitted signals $x$.};
% 矢印: Channel calc → MSE
\draw[line] (chan_calc.south) |- (mse.east);
% 矢印: PUSCH → AWGN（変換は ground-truth 後）
\draw[line] (pusch.south) -- ++(0,-5mm) -| (awgn.north);
% 矢印: AWGN → AI
\draw[line] (awgn.south) -- (ai_ce.north);
% 矢印: AI → MSE
\draw[line] (ai_ce.south) |- (mse.west);

%================== DB → Sionna 矢印 ==================
\draw[line,dashed] (fapi.east) -- ++(8mm,0) |- (pusch.west)
                    node[midway, above, font=\footnotesize]{PUSCH cfg.};
\draw[line] (fapi.north) -- ++(0,18mm) -| (pusch.north)
                    node[pos=0.2, above, font=\footnotesize]{Transmitted bits $b$.};
\draw[line] (fh.east) -- ++(10mm,0) |- ([yshift=1mm]awgn.west)
                    node[pos=0.2, left, font=\footnotesize]{Received signals $y$.};

\end{tikzpicture}}
    \caption{Training pipeline for channel frequency interpolation model.}
    \label{fig:channel_estimation_diagram}
\end{figure}

The OTA evaluation is carried out in a controlled indoor chamber with line-of-sight (LoS) configuration. To emulate interference, a programmable signal generator transmits OFDM-like noise concurrently with the PUSCH transmission. We compare our Transformer model with an in‑house 8‑layer ResNet‑based CNN baseline, which replaces the default Aerial PHY pipeline and performs channel estimation and frequency‑domain interpolation end‑to‑end.

To minimize latency and resource consumption, the Transformer model used in this task is configured with only \textbf{1 layer and 1 attention head}, in contrast to the deeper 4-layer 4-head configuration used in other tasks. This design choice reflects the relatively simpler nature of frequency interpolation and emphasizes the model’s adaptability to lightweight deployment settings. All inference measurements were conducted using \textbf{TensorRT on an NVIDIA GH200 Superchip}.

\begin{figure}[t]
    \centering
    \includegraphics[width=1\linewidth]{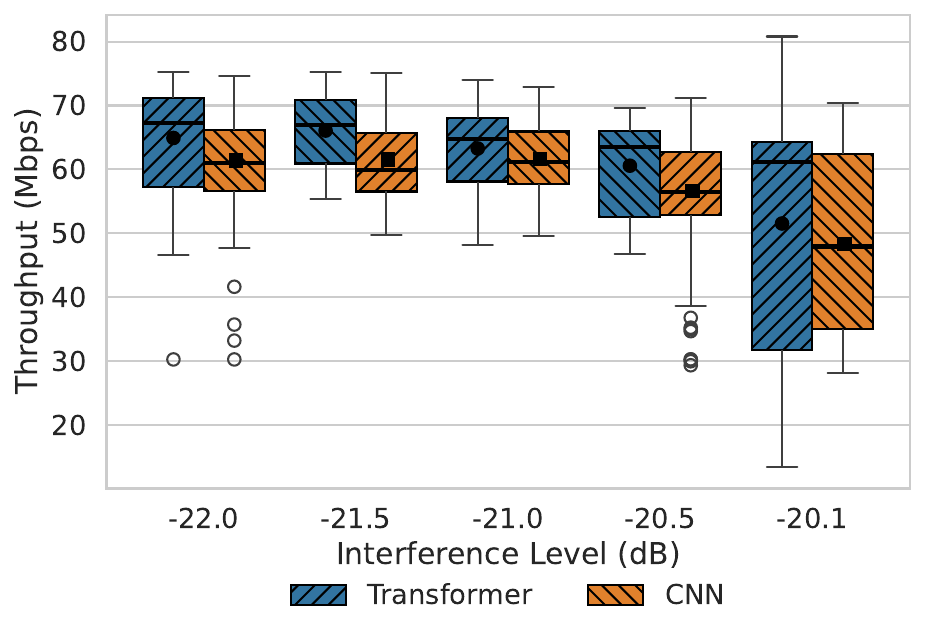}
    \caption{OTA evaluation: Uplink throughput across interference levels (Transformer: 1L1H, CNN baseline).}
    \label{fig:ota_throughput}
\end{figure}

As shown in Figure~\ref{fig:ota_throughput}, the Transformer model consistently achieves higher uplink throughput than the CNN-based baseline, across varying levels of interference. Notably, this performance gain is achieved despite using a highly compact Transformer configuration, further highlighting the effectiveness of the architecture in low-complexity settings.

While throughput serves as a key metric for communication performance, processing latency is an equally critical factor in real-world deployments, particularly for time-sensitive PUSCH chains in 5G and beyond. To assess this, we measured the average execution time per PUSCH pipeline invocation during OTA experiments—that is, the end-to-end processing time of the uplink PUSCH chain, including the model block as well as surrounding pre-/post-processing and tensor conversions.

\begin{table}[t]
  \centering
  \caption{Mean end-to-end (E2E) PUSCH receiver latency during OTA evaluation; speedup is relative to the CNN baseline.}
  \label{tab:ota_latency}
  \begin{tabular}{lcc}
    \hline
    \textbf{Model} & \textbf{E2E Latency [\(\mu\)s]} & \textbf{Speedup [\(\times\)]} \\
    \hline
    Transformer (1L1H) & 337.70 & $1.36\times$ faster \\
    CNN & 458.84 & baseline \\
    \hline
  \end{tabular}
\end{table}

As shown in Table~\ref{tab:ota_latency}, the compact Transformer model achieves a lower average PUSCH pipeline time of 337.70~$\mu$s versus 458.84~$\mu$s for the CNN baseline, corresponding to a $1.36\times$ speedup. These results indicate that, even under stringent latency and hardware constraints, the proposed Transformer architecture can reduce end-to-end uplink processing time with a compact model footprint.

\subsection{CHANNEL ESTIMATION}
\label{sec:channel_estimation}

We finally evaluate the proposed Transformer model on the task of channel estimation—a fundamental PHY-layer operation that aims to recover the full channel response across a time-frequency resource block (RB) from the received signal. This task represents a traditional signal processing problem, allowing us to validate the architectural generality of the model beyond end-to-end or pilot-based inference.

In contrast to the frequency interpolation task—which reconstructs missing subcarriers based on sparsely placed pilots—channel estimation seeks to recover the full time-frequency channel matrix, including both pilot and data resource elements. The model is trained to infer the entire complex-valued channel surface across 12 subcarriers and 14 OFDM symbols within an RB.

The input to the model is the received signal $y$ for one RB, where $y$ implicitly encodes the impact of the wireless channel, modulation, and noise. The model outputs an estimate $\hat{h}$ of the channel matrix $h$, where each complex entry is represented by two real-valued components (real and imaginary). The resulting output tensor thus has a shape of $(12 \times 14 \times 2)$ per receive antenna.

\begin{table}[t]
\centering
\caption{Simulation Parameters for Channel Estimation}
\label{tab:estimation_params}
\begin{tabular}{ll}
\hline
\textbf{Parameter} & \textbf{Value} \\
\hline
Carrier Frequency & 3.5 GHz \\
Subcarrier Spacing & 30 kHz \\
RB Size & 12 SC $\times$ 14 sym \\
Antenna Configuration & 1 UE (1 Tx), 1 BS (4 Rx) \\
Modulation Scheme & 256-QAM \\
Channel Model & CDL-C (3GPP) \\
Loss Function & Mean Squared Error (MSE) \\
Evaluation Framework & Sionna (TensorFlow) \\
\hline
\end{tabular}
\end{table}

To train the model, we adopt a supervised learning setup using synthetic data generated via the CDL-C model under varying $E_b/N_0$ conditions. The ground-truth channel $h$ is computed analytically from the simulation, and the model is optimized to minimize the MSE between $\hat{h}$ and $h$ over the full RB.

We compare our model to three baselines:
\begin{itemize}
  \item \textbf{Perfect CSI}: Uses true channel coefficients for reference (upper bound).
  \item \textbf{LS Estimation + Linear Interpolation}: Applies least squares estimation at DMRS positions, followed by 2D linear interpolation in time and frequency.
  \item \textbf{LS Estimation + Nearest Neighbor Interpolation}: Uses nearest-pilot assignment to fill non-DMRS positions.
\end{itemize}

\begin{figure}[t]
    \centering
    \includegraphics[width=1\linewidth]{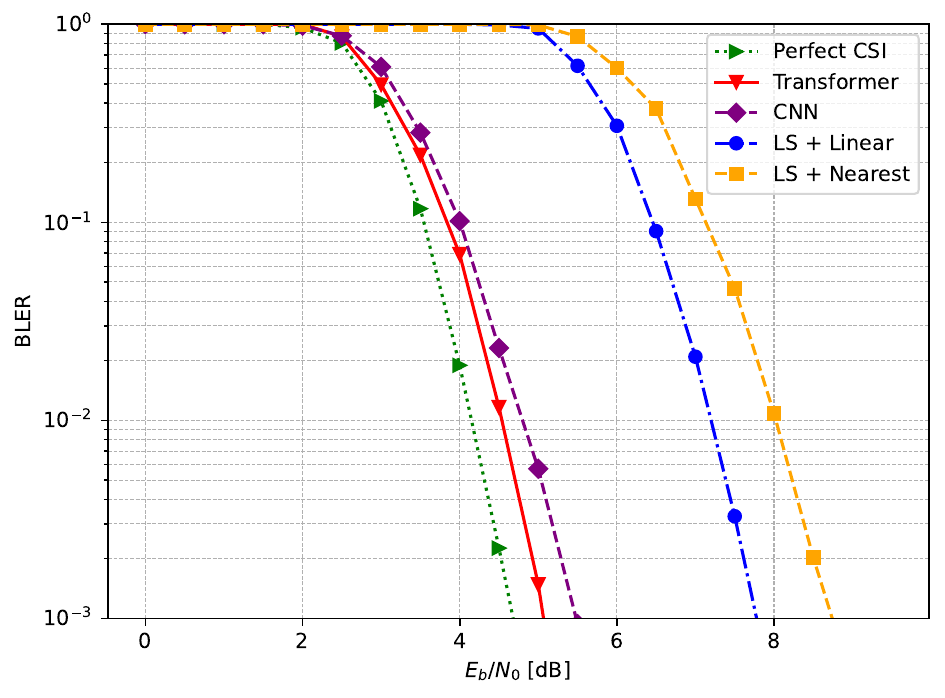}
    \caption{Simulation results: Channel estimation BLER across different methods.}
    \label{fig:estimation_sionna_comparison}
\end{figure}

Figure~\ref{fig:estimation_sionna_comparison} shows that the Transformer significantly outperforms both LS-based baselines across all $E_b/N_0$s, consistently achieving lower estimation error. At higher $E_b/N_0$s, the model's performance approaches that of the perfect CSI reference, indicating that it can effectively infer channel structure beyond pilot positions by exploiting global dependencies in the received signal.

These results demonstrate the model’s robustness and generalization capabilities in a low-$E_b/N_0$ regime, reinforcing its suitability as a general-purpose architecture for wireless signal processing tasks ranging from traditional estimation to end-to-end inference.

\section{DISCUSSION}
\label{sec:discussion}

This section analyzes key factors affecting the design and deployment of the proposed Transformer-based receiver, including architectural trade-offs, training behavior, latency constraints, and system-level integration. Through these analyses, we provide insight into the strengths and limitations of Transformer models for real-world wireless communication tasks.

\subsection{TRANSFORMER DEPTH AND ATTENTION HEAD COUNT}
Transformer depth is a primary determinant of the model’s representational capacity. In our experiments, we observe that shallow configurations—typically one to four layers—are sufficient for most tasks, especially in scenarios with limited spatial or temporal signal diversity (e.g., single-user channels with short symbol lengths). While deeper models offer modest performance gains, the associated increase in inference latency and memory footprint often outweighs the benefits under real-time constraints~\cite{Tay2022Efficient,Cai2022TRNAMC}. Similarly, expanding the number of attention heads beyond four shows little benefit and can even promote overfitting, consistent with prior observations of head redundancy in Transformers~\cite{Michel2019SixteenHeads,Voita2019Heads}. These observations suggest that wireless signal processing tasks—unlike language or vision—may benefit more from architectural simplicity and frequency-aware preprocessing than from scaling layers or attention heads.

\subsection{LATENCY-CONSTRAINED EFFICIENCY}
Meeting sub-millisecond latency budgets is a critical requirement for practical deployment in systems like O-RAN and URLLC. To this end, our architecture was explicitly designed with minimal preprocessing overhead, shallow Transformer stacks, and token-efficient input formats. Runtime profiling on an NVIDIA GH200 Superchip, as well as real-time execution in the Aerial stack, indicates that inference times remain well within acceptable bounds for GPU-equipped edge servers co-located with base stations. While our experiments use a high-end datacenter GPU, the compactness of the model makes it amenable to deployment on lower-power edge GPUs (e.g., NVIDIA L4, Orin) with further optimization through structured pruning, quantization-aware training, and hardware co-design techniques such as kernel fusion and TensorRT acceleration.

\subsection{TASK SCALABILITY AND MODULATION ADAPTIVITY}
A key design objective was to support multiple receiver tasks—such as demapping, interpolation, and estimation—within a shared Transformer backbone. This was achieved by adjusting only the final projection layer and loss function, depending on whether the output is soft bits, channel state matrices, or regression targets. The model also exhibits modulation adaptivity: when trained with high-order constellations like 256QAM, the same model can generalize to lower-order formats like QPSK by selectively decoding bit positions aligned with Gray-coded signal space. This enables deployment of a single model across adaptive modulation and coding (AMC) schemes, reducing maintenance and storage overhead.

\subsection{SYSTEM INTEGRATION AND REAL-WORLD DEPLOYMENT}
Beyond simulation, the model was validated in the OAI+Aerial testbed, integrated with 3GPP-compliant baseband stacks via OAI and Aerial. This demonstrates feasibility in real-time, over-the-air operation under controlled but realistic RF environments. However, full-scale deployment in commercial networks introduces additional requirements: KPI traceability, scheduling robustness, support for mobility and handovers, and resilience under interference. Additionally, system-level pipelines for managing ML lifecycle—such as model catalog versioning, region-aware fine-tuning, and over-the-air updates—will become increasingly important as learning-based modules proliferate in RAN infrastructure.

\vspace{0.5em}
Overall, our analysis underscores that Transformers can serve as a unified processing block for next-generation receivers, provided that architectural decisions are grounded in domain-specific constraints and deployment requirements. These findings motivate further research into hardware-aware design, modularity, and lifecycle management of ML-enabled radio access networks.

\section{FUTURE WORK}
\label{sec:future_work}
Building on the findings of this study, we outline both immediate next steps and broader research directions aimed at further enhancing the performance, efficiency, and deployability of Transformer-based architectures in wireless communications. These efforts are intended to address current limitations while expanding applicability to more diverse and complex scenarios in next-generation wireless systems.

\subsection{IMMEDIATE NEXT STEPS}

\subsubsection{Extending Beyond Layer-1 Tasks}
Beyond the Layer-1 tasks explored in this work, we will evaluate the proposed architecture on additional tasks beyond L1, including cross-layer (e.g., MAC) settings, to assess its generality and identify task-specific adaptations needed for optimal performance.

\subsubsection{Improved Positional Embeddings}
We plan to investigate alternative positional encoding schemes better suited for structured time–frequency–antenna domains, with the goal of improving performance under diverse channel and pilot configurations.

\subsubsection{Lightweight Inference via Model Optimization}
To further reduce inference latency and energy consumption, we will explore quantization, pruning, and early-exit mechanisms. These methods will be assessed in terms of accuracy–efficiency trade-offs and their impact on real-time deployability under strict latency budgets.

\subsubsection{Cross-Attention-Based Information Compression}
We will explore attention-based mechanisms for selectively summarizing less critical input features into compact representations, enabling reduced computational overhead and lower power consumption when operating across multiple tasks.

\subsubsection{Task-Integrated Efficiency Evaluation}
The above techniques will be jointly evaluated across multiple tasks to quantify end-to-end gains in runtime and energy efficiency, and to inform co-design strategies for system integration.

\subsubsection{Robustness to Channel Dynamics}
Our evaluation to date focuses on low-to-moderate mobility, leaving robustness under high-Doppler, rapid Rayleigh fading, and other fast-varying conditions untested. 
In the current resource grid (RG) input setting, such variations may be captured within a single RG snapshot; however, scenarios involving multiple consecutive RGs or longer-term dependencies—such as cross-slot scheduling or HARQ processes—could expose limitations of the Transformer's stateless design and reliance on positional encodings.
Potential directions include incorporating lightweight mechanisms to maintain temporal context across RGs or adapting positional representations to remain stable under mobility-induced distortions.

\subsection{BROADER RESEARCH DIRECTIONS}

\subsubsection{Multimodal and Cross-Layer Learning}
Integrating diverse inputs—CSI, SNR, mobility metrics, and scheduling metadata—into a unified learning pipeline could enable cross-layer optimization strategies and improve robustness in real-world deployments.

\subsubsection{Model Interpretability and Debuggability}
We will examine whether attention concentrates on pilot positions and how such patterns vary with interference/mobility, and assess if these signals can guide adaptive pilot design or targeted retraining triggers.

\subsubsection{Standardization and System-Level Validation}
Extending evaluation to full-stack, 3GPP-compliant scenarios with real traffic profiles, and developing lifecycle management frameworks for distributed RAN deployments, are critical for operational adoption.

By combining targeted short-term efforts with broader, exploratory research, we aim to advance learning-driven physical layer techniques toward practical, efficient, and trustworthy deployment in future wireless systems.

\section{CONCLUSION}
\label{sec:conclusion}

This paper presented a unified, latency-aware Transformer-based architecture that operates directly on the resource grid and can substitute several PHY blocks—channel estimation, equalization, and demapping—using a compact self-attention backbone. The design targets real-time feasibility under strict latency budgets while remaining adaptable across tasks.

We validated the approach on three representative tasks: end-to-end receiver, channel frequency interpolation, and channel estimation. In the end-to-end receiver setting, the model processed the full pipeline from pilot observations to bit-level decisions and maintained strong performance across modulation orders and user counts. For channel frequency interpolation, the model was integrated into a 3GPP-compliant OAI+Aerial stack and evaluated over the air, confirming real-world feasibility. In channel estimation, the model reconstructed channels from sparse pilots and outperformed classical LS with linear and nearest-neighbor interpolation.

Across tasks, a single backbone with lightweight task-specific heads delivered robust accuracy and inference efficiency, requiring only minimal changes to adapt output formats and objectives.

Open challenges remain in improving generalization under rapidly varying operating conditions (e.g., high Doppler and bursty interference), strengthening efficiency within tight latency and energy budgets via compression and early-exit strategies, refining positional representations for diverse pilot layouts, and developing interpretability and monitoring tools that expose how pilot information is exploited. In addition, standardized KPI-driven full-stack evaluations and lifecycle-aware deployment practices are needed to bridge prototypes and production.

In summary, our work highlights the potential of a task-adaptive, attention-based architecture to replace hand-engineered signal processing chains in next-generation communication systems. By applying the same compact Transformer backbone across diverse PHY-layer tasks and validating its performance in both simulation and OTA conditions, we demonstrate a practical path toward learning-driven, software-defined physical layers for 5G and beyond.

\bibliographystyle{IEEEtran}
%\bibliography{references}

\end{document}